%% file: ttsusy.tex
\documentstyle[epsf]{article}
\epsfverbosetrue
\def\PRL #1 #2 #3 {Phys.~Rev.~Lett.~{\bf #1}, #2 (#3)}
\def\PRD #1 #2 #3 {Phys.~Rev.~D~{\bf #1}, #2 (#3)}
\def\PLB #1 #2 #3 {Phys.~Lett.~{\bf B#1}, #2 (#3)}
\def\NPB #1 #2 #3 {Nucl.~Phys.~{\bf B#1}, #2 (#3)}
\def\NPBPS #1 #2 #3 {Nucl.~Phys.~B~(Proc.~Suppl.)~{\bf #1B}, #2 (#3)}
\def\RMP #1 #2 #3 {Rev.~Mod.~Phys.~{\bf #1}, #2 (#3)}
\def\PREP #1 #2 #3 {Phys.~Rep.~{\bf #1}, #2 (#3)}

\newcommand{\beq}{\begin{equation}}
\newcommand{\eeq}{\end{equation}}

\newcommand{\sq}[1]{\mbox{$\widetilde{#1}$}}

\def\xmt {\mbox{$m^2_t$}}
\def\xmqu {\mbox{$m^2_q$}}

\def\xmg {\mbox{$m^2_{\widetilde{g}}$}}
\def\xmq {\mbox{$m^2_{\widetilde{q}}$}}
\def\xmQ {\mbox{$m^2_{\widetilde{Q}}$}}

\def\xth {\mbox{$\theta_{\widetilde{t}}$}}
\def\xthq {\mbox{$\theta_{\widetilde{q}}$}}
\def\xthQ {\mbox{$\theta_{\widetilde{Q}}$}}
\def\xmst {\mbox{$m^2_{\widetilde{t}}$}}
\def\mmt {\mbox{$m_t$}}
\def\mmqu {\mbox{$m_q$}}
\def\mma {\mbox{$m_{\widetilde{t}_1}$}}
\def\mmb {\mbox{$m_{\widetilde{t}_2}$}}
\def\mmg {\mbox{$m_{\widetilde{g}}$}}

\def\mmq {\mbox{$m_{\widetilde{q}}$}}
\def\mmQ {\mbox{$m_{\widetilde{Q}}$}}

\def\mmst {\mbox{$m_{\widetilde{t}}$}}
\def\acbd {\mbox{$a\,c+b\,d$}}
\def\asbs {\mbox{$a^2+b^2$}}
\def\csds {\mbox{$c^2+d^2$}}

\begin{document}
\begin{titlepage}
 
\medskip
\rightline{hep-ph/9611302}
\rightline{ILL-(TH)-96-13}
\rightline{November 1996}
\bigskip\bigskip\bigskip

\begin{center}
{\Large\bf Supersymmetric QCD correction to top-quark}\\
{\Large\bf  production at the Tevatron}\\
\bigskip\bigskip\bigskip\bigskip
{\large{\bf Zack Sullivan}}\\
\medskip Department of Physics\\
University of Illinois\\
1110 West Green Street\\
Urbana, IL  61801\\ 
\end{center} 
\bigskip\bigskip\bigskip\bigskip

\begin{abstract}
We calculate the supersymmetric QCD correction to top-quark production at the
Fermilab Tevatron, allowing for arbitrary \mbox{left-right} mixing of the
squarks.  We find that the correction is significant for several combinations of
gluino and squark masses, e.g. $+33\%$ for $\mmg=200$~GeV, $\mmst=\mmq=75$~GeV.
\end{abstract}

\addtolength{\baselineskip}{9pt}

\end{titlepage}

\begin{center}
\section{Introduction}
\end{center}

\indent The discovery of the top quark \cite{TOPD} provides a unique opportunity
to search for effects beyond the Standard Model.  The top quark mass $\mmt=
175\pm 6$~GeV has been measured to 3.5\%, and the cross section has been
measured to $\approx25\%$ \cite{TOPM}.  With the copious production of top
quarks in Run II of the Fermilab Tevatron and future upgrades, the cross section
will be measured to within 6\% with 10~fb$^{-1}$ of data \cite{TEV2000}.
Comparison of the theoretical cross section to that measured will test the
Standard Model and may indicate the presence of new physics.

  Supersymmetry is a promising candidate for new physics.  Currently, only lower
bounds on the masses of the superpartners have been set.  Barring discovery,
direct searches for SUSY will eliminate a small range of parameter space, since
these searches depend strongly on the modelling of the decays of the
supersymmetric particles.  In contrast, some effects of virtual supersymmetry
are less model dependent, thus extending the reach of experiment.  If virtual
SUSY effects are found to be large enough, an indirect search may provide the
first sign of supersymmetry.  In this paper we calculate the supersymmetric QCD
correction to $t\bar t$ production at the Fermilab Tevatron.

Direct searches for SUSY are generally performed separately for top squarks, the
light quark superpartners, and gluinos.  This is motivated by minimal
supergravity models which argue that all scalar particles acquire a mass on the
order of the SUSY breaking scale \cite{NILLESAR}.  A heavy top quark loop
dominates the running of the masses to low energies, forcing the mass of the two
top squarks below that of the rest of the squarks.  Additionally, mixing of the
\mbox{left-right} weak eigenstates of the top squarks may result in the top
squark $\sq{t}_1$ becoming the lightest squark \cite{ELLISMIX}.  The lightest
top squark has a mass limit of $\mma > 47$~GeV \cite{ALSUSY}.  The mass limit of
the gluino is $\mmg > 154$~GeV, when the light quark superpartners are assumed
to be heavier than the gluino \cite{CDFSG}.  Current experimental limits are
extracted assuming specific values of the SUSY parameters, and may be relaxed
\cite{DATTA}.  Other regions of parameter space have been eliminated
\cite{OSUSY,D0ST}, but these limits are generally model dependent.  Exhaustive
direct searches will reach 300~GeV for gluinos and 100~GeV for top squarks with
10~fb$^{-1}$ of data at the Tevatron \cite{TEV2000}.  Until the advent of the
CERN Large Hadron Collider, the presence of heavier SUSY particles will only be
suggested by their effects on Standard Model processes.
	
The NLO QCD cross section for $t\bar t$ production with resummed gluon emission
at a $\sqrt{S} = 2$~TeV $p\bar p$ collider has been calculated \cite{CROSSTH}.
The dominant mechanism of top-quark production at the Tevatron is $q\bar q$
annihilation.  It is expected that the dominant SUSY contribution to top
production will be in the form of QCD corrections to this process.  We consider
the SUSY correction to the cross section as a correction to the dominant
process as shown in Fig.~1.
The calculation of the SUSY correction to top-quark production is different from
typical SUSY calculations in that the number of assumptions about supersymmetry
necessary to predict phenomenologically interesting results is minimal.  It is
assumed that R-parity is conserved so that the interaction terms in the
Lagrangian are the simple supersymmetrization of the Standard Model
interactions.  No assumptions about the mechanism of SUSY breaking or of
unification are required.  In a strong-interaction process, the correction
depends only on the observed masses of the gluino and squarks, and the mixing
angle that relates the squark mass eigenstates to their interaction eigenstates.
For the purpose of this calculation, and in order to cover the greatest range of
models, we treat top squarks $\sq{t}$ separately from the light-quark
superpartners $\sq{q}$.  We present analytic and numerical results for
degenerate squark masses, and for the case where the top squarks are light
compared to the light-quark superpartners, the `heavy squarks'.  Results for
$\mmst,\mmq >50$~GeV, $\mmg > 150$~GeV, and arbitrary \mbox{left-right} mixing
of the squarks are presented.

The SUSY QCD correction to top production in $e^+e^-$ annihilation has been
studied in Ref.~\cite{DJOUADI}.  The correction in $p\bar p$ annihilation has
been presented in Refs.~\cite{LIHU,KIMLOP,ALAM} for the case of degenerate
squark masses.  The calculations of Refs.~\cite{LIHU,KIMLOP} neglect the
contribution of the vacuum polarization and the crossed-box diagram, which
arises because the gluino is a Majorana particle.  In addition,
Ref.~\cite{KIMLOP} assumes that the box contribution may be ignored.  We find
that these contributions are numerically significant, as demonstarted in
Sec.~III~A.  Our results are numerically comparable to Ref.~\cite{ALAM}, however
there is an important sign discrepancy between the two box terms that we discuss
in Sec.~II.  There also appears to be a misprint in the analytic results of that
paper that we describe at the end of our Appendix.  We provide a complete
calculation of the SUSY correction to the cross section for arbitrary masses and
top-squark mixing, and discuss the phenomenological significance of the result.
In addition, we discuss the tenability of finding SUSY thresholds in $t\bar t$
invariant mass distributions.  Finally we address the issue of parity violation
in a supersymmetric strong force.

This paper is organized as follows.  In Section~II, we present the analytic form
of the ${\cal O}(\alpha_s)$ SUSY QCD correction to the $p\bar p \to t\bar t$
cross section.  In Section~III~A, we remark on the relative size of the terms in
the correction.  We present numerical results for the correction to the $p\bar p
\to t\bar t$ cross section in Sec.~III~B.  In Section~III~C, we show $t\bar t$
invariant mass distributions for several choices of gluino mass.  We discuss the
size of the strong force parity-violating \mbox{left-right} asymmetry in
Sec.~III~D.  Conclusions are presented in Sec.~IV.  We present analytic
expressions for the vacuum, vertex and box terms in the Appendix.

\begin{center}
\section{Analytic Supersymmetric QCD Correction}
\end{center}

\indent The one-loop supersymmetric QCD contribution to the $q\bar q \to t\bar
t$ cross section at leading order in $\alpha_{s}$ is attributed to the cross
term in the matrix element between the tree level diagram and the one-loop
diagrams presented in Fig.~1.  The general form of the vertex
corrections, consistent with current conservation, is
\begin{eqnarray}
\lefteqn{i\bar u(p_1)\Gamma^{\mu,A} v(p_2) = -i\,g_s \Biggl[\bar u(p_1) T^A
\gamma^\mu v(p_2)\Biggr. }       \\
 & & \mbox{} \Biggl.  - \frac{\alpha_{s}}{4\pi}\;\bar u(p_1) T^A\left[
V\gamma^\mu + S(p_{1}^{\mu} - p_{2}^{\mu})/\mmqu +A (\gamma^{\mu}q^2 - 2\mmqu
q^{\mu})\gamma_5\right] v(p_2) \Biggr] \;, \nonumber
\end{eqnarray}
where $p_{1}$ and $p_{2}$ are the momenta of the quark and antiquark, $q =
p_{1} + p_{2}$, $T^{A}$ is a $SU(3)$ generator, and $V$, $S$, and $A$ are the
vector, scalar, and anapole form factors, respectively.  The analytic forms of
$V$, $S$, $A$, the gluon vacuum polarization $\Pi$, and the corrections due to
the box and crossed-box diagrams, $B$ and $C$, are given in an appendix.  The
anapole term $A$ does not contribute to the total cross section at this order in
the expansion.  It is used in Sec.~III~D, however, in determining the
parity-violating \mbox{left-right} asymmetry due to the squark mixing.  The
Dirac algebra and loop integrals were evaluated using dimensional
regularization.  The analytic cross section was derived in the $\overline{MS}$
renormalization scheme.  The Feynman rules for the SUSY verticies were derived
from Ref.~\cite{HABERK} for the physically-relevant mass eigenstates of the
squarks rather than the interaction eigenstates.  Mixing of the squarks is
therefore explicit and parameterized by mixing angles $\xth$ and $\xthq$ for the
stops and light-quark superpartners, respectively:
\begin{eqnarray}
\left(\begin{array}{c} \sq{t}_1 \\ \sq{t}_2 \end{array} \right)&=&
\left(\begin{array}{rr} \cos \xth & \sin \xth \\
-\sin \xth & \cos \xth \end{array} \right)
\left(\begin{array}{c} \sq{t}_L \\ \sq{t}_R \end{array} \right) \\
\left(\begin{array}{c} \sq{q}_1 \\ \sq{q}_2 \end{array} \right) &=&
\left(\begin{array}{rr} \cos \xthq & \sin \xthq \\
-\sin \xthq & \cos \xthq \end{array} \right)
\left(\begin{array}{c} \sq{q}_L \\ \sq{q}_R \end{array} \right) \;. \nonumber
\end{eqnarray}

The spin- and color-averaged parton-level differential cross section is given by
\begin{equation}
\frac{d\widehat{\sigma}}{dz} = \frac{\beta}{32\pi\widehat{s}} |M|^2 \;,
\end{equation}
where $z$ is the cosine of the angle between the incoming quark and the top
quark, $\beta = \sqrt{1-4\xmt/\widehat{s}}$, and $\sqrt{\widehat{s}}$ is the
parton center-of-momentum energy.  The Born matrix element squared is given by
\begin{equation}
|M_0|^2 = \frac{32\pi^2{\alpha_s}^2}{9}[2 - \beta^2(1-z^2)] \;.
\end{equation}
Integrating over $-1\leq z\leq 1$ readily yields the Born-level cross section
\begin{equation}
\widehat{\sigma}_0 = \frac{4\pi\alpha^2_s\beta}{9\widehat{s}} (1 - \beta^2/3) \;.
\end{equation}

The correction arises from the cross term in the square of the amplitude.
This correction is the sum of the terms:
\begin{eqnarray}
2{\rm Re}[M^\dagger_0M_{\Pi}] & = & - \frac{\alpha_s}{2\pi} |M_0|^2{\rm Re}
[\Pi(\widehat{s}) - \Pi(0)] \label{MATRIX} \\
2{\rm Re}[M^\dagger_0M_V] & = & - \frac{\alpha_s}{2\pi}|M_0|^2{\rm Re}[V]
\nonumber \\
2{\rm Re}[M^\dagger_0M_S] & = & \frac{32\pi\beta^2\alpha^3_s}{9}(1 - z^2){\rm
Re}[S] \nonumber \\
2{\rm Re}[M^\dagger_0M_{BOX}] & = & \frac{32\pi\alpha^3_s}{9\widehat{s}}{\rm Re}
\left[\frac{7}{3}B + \frac{2}{3}C\right] \;. \nonumber
\end{eqnarray}

We renormalize the vacuum polarization correction so that it corresponds to the
known value of $\alpha_s$ in the $\overline{MS}$ scheme at low energy.  The
integration over phase space is trivial except for the box and crossed-box
matrix elements, $B$ and $C$, which depend implicitly on $z$.  The relative sign
between the box and crossed-box terms should be noted. The color factor
associated with $C$ is $-2/3$.  However, Fermi statistics, and the proper
ordering of the Dirac indicies of the gluino fields in the amplitude, provide a
non-trivial relative sign difference between the two diagrams.  The net result
is that the two contributions constructively interfere.  This disagrees with the
calculation of Ref.~\cite{ALAM}, which claims that the terms destructively
interfere.

The total cross section for top production in $p\bar p$ annihilation is obtained
by convolving the parton cross section for annihilation into a $t\bar t$ final
state with the parton distribution functions of the proton and antiproton.  The
integral may be parameterized as
\begin{equation}
\sigma = \int_{4m_{t}^{2}/S}^{1}d\tau\,\widehat{\sigma}(\tau 
S)\int_{\ln(\tau)/2}^{-\ln(\tau)/2}d\eta\,P(\sqrt{\tau}e^{\eta},
\sqrt{\widehat{s}}\,)\,\overline{P}(\sqrt{\tau}e^{\eta},\sqrt{\widehat{s}}\,)
\;,
\end{equation}
where $\sqrt{S} = 2$~TeV, $\tau = \widehat{s}/S$ and $P(x_1,\mu)$,
$\overline{P}(x_2,\mu)$ are the proton and antiproton parton distribution
functions (PDF's).

In the following section, numerical results are presented for a top quark of
mass $\mmt = 175$~GeV.  Analytic expressions were reduced to scalar n-point
integrals \cite{PASSARINOV} and evaluated with the aid of the code FF \cite{FF}
in order to ensure numerical stability.  For those cases that FF does not
handle, the analytic solutions to the integrals were substituted.  The integrals
were evaluated using both the MRS(A$^\prime$) \cite{MRSAP} and CTEQ3M
\cite{CTEQ} PDF's.  The coupling $\alpha_{s}$ was evaluated as in the PDF's in
order to be consistent.  Nearly identical results were obtained using both sets,
therefore, only the results obtained using the MRS(A$^\prime$) PDF's are
presented.

\begin{center}
\section{Numerical Results}

\subsection{Relative Size of the Correction Terms}
\end{center}

\indent In Figure~2 we show the correction to the total cross
section as a function of common squark mass $\mmQ \equiv \mmq =\mmst$, for $\mmg
= 200$~GeV.  The contribution of the vacuum, vector, scalar, and box terms are
shown separately.  The total correction is also shown for comparison.  The box
diagrams give the largest contribution to the cross section for $\mmQ<110$~GeV,
and are significant for $\mmQ<400$~GeV.  This invalidates the assumption of
Ref.~\cite{KIMLOP} that the box terms may be neglected over the range of masses
they investigated.  Similarly it contradicts the conclusion of Ref.~\cite{ALAM}
that the contribution of the box terms is small.  The vacuum correction, that
was ignored in Refs.~\cite{LIHU,KIMLOP}, also plays an important role.  The
gluino loops in the vacuum polarization give a constant negative correction when
the squarks decouple.  When $\mmQ=1$~TeV, the correction is seen to come almost
entirely from the vacuum polarization.  The contribution of the scalar term $S$
is negligible.  It first appears at this order in the final state correction,
and is suppressed relative to the other terms by a power of the top-quark mass.
The decoupling of the vector, scalar, and box terms is evident in
Fig.~2, as the corrections decrease when the squark mass increases.

\begin{center}
\subsection{$t\bar t$ Cross Section}
\end{center}

\indent The correction to the $p\bar p \to t\bar t$ cross section is shown in
Fig.~3 as a function of the gluino mass for a wide range of degenerate
squark masses $\mmQ$, where $\mmQ \equiv \mmq = \mmst$.  As expected from
decoupling, the magnitude of the correction decreases as the squark mass
increases.  Squarks of mass 50~GeV set the range of the correction from
$-11.8\%$ for a gluino of 150~GeV to $+44\%$ for a gluino of 200~GeV.  The
correction changes sign as $\mmg$ approaches $\mmt$.  The correction changes
rapidly as the threshold for gluino production moves through the top-quark
threshold.  Note that the correction is nearly independent of gluino mass when
$\mmg > 700$~GeV.  In this region, the correction is entirely dominated by the
squark vacuum terms and, to a lesser extent, the box terms.

In Figure~4 we show the correction to the total cross section as a
function of degenerate squark mass $\mmQ \equiv \mmq = \mmst$, for several
gluino masses.  Once $\mmQ > 400$~GeV, the correction becomes small and the
squarks effectively decouple.  In this region, the correction is dominated by
the gluino vacuum terms.  In Figs.~3 and 4 there is a large
jump in the cross section when $\xmt = \xmst + \xmg$.  In Figure 4, the
correction jumps from $+6.5\%$ to $-9.3\%$ for $\mmg = 150$~GeV, $\mmst =
90.1$~GeV.  This corresponds to a discontinuity in the real part of the $C_0$
scalar loop-integral in the final-state vertex correction \cite{THOOFTV}.  Such
a discontinuity arises when the anomalous threshold crosses the real threshold
for superpartner production in the complex $s$-plane \cite{Itzykson}.

The largest correction occurs when $\mmg = 200$~GeV.  This mass is used in
Fig.~5 to show the correction as a function of heavy-squark mass
$\mmq$, for a variety of top-squark masses.  This figure demonstrates that the
correction is mostly influenced by the mass of the top squark.  For example, the
correction is 21\% for $\mmst = 50$~GeV, and $\mmq = 300$~GeV; whereas the
correction is 16\% for $\mmst = 300$~GeV, and $\mmq = 50$~GeV.  Even if the
heavy squarks decouple, the correction remains significant as long as $\mmst <
150$~GeV.

In general, the left and right eigenstates of the squarks receive different
corrections to their masses.  This causes the mass of $\sq{Q}_R$ to be less than
the mass of $\sq{Q}_L$.  Top-squark masses are more effected by renormalization
group running than the heavy-squark masses, because of the direct coupling of
the top to the stops.  Many analyses assume that the only light squark is
$\sq{t}_1$, and look for top quarks decaying into them \cite{D0ST}.  In
Figure~6 we show the ratio of the correction at
$\Delta\mmst=(\mmb-\mma)$ to the correction at a common top-squark mass
$\Delta\mmst=0$, for $\mmg = 200$~GeV and $\mmq = 300$~GeV.  The ratio does not
change by more than 2\% for different values of $\mmq$.  We present three mixing
angles, $\xth = 45^{\circ}$, $90^{\circ}$, and $135^{\circ}$ that define the
extremes of the mixing dependence of the correction.  The form of the correction
is $a+b\sin{(2\xth)}$, thus the contribution of any mixing angle may be
interpolated from the curves shown, where $\xth=90^{\circ}$ is the central
value.  Note that if $\xth=135^{\circ}$, then the correction is nearly
independent of $\mmb$; whereas if $\xth=90^{\circ}$, where the mass eigenstates
are the interaction eigenstates, the correction is roughly split between the two
top squarks.  To evaluate the correction for non-degenerate top-squark masses
and top-squark mixing, multiply the ratio from Fig.~6 by the
correction from Figs.~4 or 5.  For example, the correction to
top production is $7.7\pm 0.1\%$, when $\mma=100$~GeV, $\mmb=400$~GeV,
$\mmg=200$~GeV, $\mmq=400$~GeV, and $\xth=90^{\circ}$.

\begin{center}
\subsection{$t\bar t$ Invariant Mass Distributions}
\end{center}

\indent Since total cross section measurements are difficult to normalize, it is
advantageous to look for deviations from the line-shapes predicted by the
Standard Model.  A sampling of the invariant mass of $t\bar t$ events provides
another avenue to search for supersymmetry.  In Figure~7 we show the
total differential cross section as a function of $t\bar t$~invariant mass
$M_{t\bar t}$, for gluinos of mass $\mmg = 150$, 175, 200, and 225~GeV.  Several
choices of degenerate squark mass $\mmQ \equiv \mmq =\mmst$, are presented.  By
looking for an excess in the invariant mass distribution, a gluino of mass
between 175~GeV and 225~GeV may be observable.

There are two types of enhancement to the cross section that appear in
Fig.~7.  If $\mmg\approx\mmt$, the maximum of the invariant mass
distribution is shifted toward the common threshold.  This would also produce a
steeper top-quark threshold region in the data.  A singularity at the threshold
for gluino pair production causes a cusp at $2\mmg$.  The largest cusp occurs
when $\mmg=200$~GeV, and $\mmQ=50$~GeV.  The amplitude of the cusp is 112\% of
the Standard Model differential cross section at this point.  Despite the large
normalization, the cusp will sit on a large continuum background.  If we assume
purely statistical errors, this cusp would appear at the $3\sigma$ level with
3~fb$^{-1}$ of integrated luminosity.  For $\mmg\approx200$~GeV, the correction
is most apparent for $\mmQ<150$~GeV.  If $\mmg>225$~GeV, then even with light
squarks, the correction will be difficult to observe.

\begin{center}
\subsection{Strong Force Parity Violation}
\end{center}

\indent In the Standard Model, the top quark decays before its spin
flips~\cite{BIGI}.  The helicity of the top quark is reflected in the angular
distribution of the decay products of the $W$~boson in $t\to bW\to b\ell^{+}\nu$
and $t\to bW\to b\bar d u$ decays. (See Ref.~\cite{JEZ} for a detailed account
of the analyzing power of these decays.)  The $gt\sq{t}$ interaction term in the
SUSY Lagrangian treats left and right-handed top squarks differently.  This
leads to the interesting possibility of searching for parity violation in strong
force interactions by analyzing the decay products in top-quark production.

An asymmetry in the number of left and right-handed top quarks arises in the
production cross section when the top squarks have different masses.  This
asymmetry is given by
\begin{equation}
\Delta\widehat{\sigma}_A = \widehat{\sigma}_L - \widehat{\sigma}_R =
\frac{2\beta^2\alpha^3_s} {27\widehat{s}}{\rm Re}[A] \;,
\end{equation}
where $\sigma_L$, $\sigma_R$ are the cross sections for the left and right
helicities of the top quark.  The measured \mbox{left-right} asymmetry $A_{LR}$
is the ratio of the integrated $\Delta\widehat{\sigma}_A$ to the total measured
cross section
\begin{equation}
A_{LR} = \frac{\Delta\sigma_A}{\sigma_{TOT}} = \frac{n_L-n_R}{n_L+n_R} \;,
\end{equation}
where $n_L$, $n_R$ are the number of left and right-handed top quarks
respectively.  Unfortunately, we find that $A_{LR}$ is always less than 1\% for
any choice of the SUSY parameters.  Therefore, if supersymmetric parity
violation in the strong force exists, it will be very difficult to measure.

\begin{center}
\section{Conclusions}
\end{center}

\indent The supersymmetric QCD correction to the top-quark cross section, as
measured at the Tevatron, has been calculated.  We present analytic results for
a minimal supersymmetric model that depends only on the masses of the
superpartners and their mixing.  We obtain numerical results for the total
correction for all masses $\mmg>150$~GeV, $\mmq>50$~GeV and $\mmst>50$~GeV.  The
correction is found to be large for gluino masses near 200~GeV.  The correction
is greater than $+10\%$ for $\mmg = 200$~GeV and $\mmq=\mmst<190$~GeV.  If light
top squarks $\mmst<150$~GeV exist, then the correction should be observable with
10~fb$^{-1}$ at the Tevatron for $\mmg<400$~GeV, even if the heavy squarks
decouple.  If all of the squarks remain light, then the correction is
significant even if the gluinos decouple.  When considering a mass splitting
between the top squarks, the mixing angle $\xth$ plays an important role.  If
$\xth$ is near $45^{\circ}$, or $135^{\circ}$, then the correction is almost
entirely dependent on the mass of only one of the top squarks.

Should the gluino mass turn out to be near the current experimental limits, a
gluino-pair threshold may be found near the top-quark production threshold.  The
advantage of looking for a cusp in the $t\bar t$ invariant mass distribution, is
that the normalization of the top-quark cross section is not necessarily a
limiting factor.  Detector resolution effects and smearings will make this
search very challenging.  It is reasonable to expect that at least 10~fb$^{-1}$
of integrated luminosity would be required to find a cusp for the best case of
$\mmg\approx200$~GeV, and $\mmQ<150$~GeV.  Virtual SUSY thresholds are common in
quark production \cite{ELLISR}.  A full detector-based analysis of these
threshold regions would help determine the experimental significance of our
results.

Parity violation in a purely strong force interaction arises in a supersymmetric
Standard Model because the left- and right-handed top squarks interact
differently.  As long as the top-squark masses are different, an asymmetry in
the number of left and right-handed top quarks will arise.  Unfortunately, the
effect is less than 1\%, and will be very difficult to measure.

\begin{center}
\section*{Acknowledgements}
\end{center}

\indent The author is grateful to S.~Willenbrock for his helpful comments and
for suggesting this study, and for conversations with T.~Stelzer.  I gratefully
acknowledge the support of a GAANN fellowship, under grant number DE-P200A40532,
from the U.~S.~Department of Education.

\begin{center}
\section*{Appendix}
\end{center}

\indent The form factors for the one-loop matrix elements in Eqn.~\ref{MATRIX}
of the supersymmetric QCD correction to top-quark production are given below.
The integrals are written in terms of n-point integrals \cite{PASSARINOV}, in
the notation of FF~\cite{FF}.  For each appearance of a heavy squark $\sq{q}$,
or top squark $\sq{t}$, the term should be summed with $\sq{q}_1$, or $\sq{t}_1$
first, and then $\pm\sq{q}_2$, or $\pm\sq{t}_2$ as indicated.  The vacuum
polarization is separated into terms proportional to the top squarks $\sq{t}$
and the heavy squarks $\sq{q}$.
\begin{eqnarray*}
\Pi &=& \frac{1}{s}\biggl[ (4\xmg+2s)B_{0}(\xmg,\xmg,s)-4A_{0}(\xmg)+
\frac{2}{3}[A_{0}(\xmst)+5A_{0}(\xmq)]\biggr. \\
& & \mbox{} \biggl. +\frac{1}{6}(s-4\xmst)B_{0}(\xmst,\xmst,s)
+\frac{5}{6}(s-4\xmq)B_{0}(\xmq,\xmq,s) \biggr. \\ & &
\mbox{} \biggl. -4s\Delta +4\xmg +\frac{2}{3}(s-\xmst -5\xmq)\biggr]
\end{eqnarray*}

The initial- and final-state vertex correction form factors have the same
functional form.  For the initial state, $\mmqu=0$, and $\mmQ=\mmq$.  For the
final state, $\mmqu=\mmt$, and $\mmQ=\mmst$.  The two squarks $\sq{Q}_1$ and
$\pm\sq{Q}_2$ are summed as before.  Arbitrary mixing is allowed for both the
top squarks and the heavy squarks.  The 3-point integrals have the form
\[
C(m_1^2,m_2^2,m_3^2) = C(m_1^2,m_2^2,m_3^2,\xmqu,\xmqu,s) \;.
\]
\begin{eqnarray*}
V &=& -\frac{3}{2}\biggl[ \xmqu C_{21}(\xmg,\xmQ,\xmg)+\xmqu
C_{22}(\xmg,\xmQ,\xmg)+(s-2\xmqu)C_{23}(\xmg,\xmQ,\xmg) \biggr. \\ & & \mbox{}
+\biggl. 2C_{24}(\xmg,\xmQ,\xmg)-1+2\xmqu C_{11}(\xmg,\xmQ,\xmg)
+(s-2\xmqu)C_{12}(\xmg,\xmQ,\xmg) \biggr. \\ & & \mbox{} -
\biggl. \xmg C_{0}(\xmg,\xmQ,\xmg) \pm
2\mmqu\mmg\sin{(2\xthQ)}C_{0}(\xmg,\xmQ,\xmg) \biggr] +
\frac{1}{3}C_{24}(\xmQ,\xmg,\xmQ) \\ & & \mbox{} +\frac{4}{3}\biggl[
-B_{1}(\xmQ,\xmg,\xmqu)+(\xmqu+\xmg-\xmQ)B^{\prime}_{0}(\xmQ,\xmg,\xmqu)
\mp 2\mmqu\mmg\sin{(2\xthQ)}B^{\prime}_{0}(\xmQ,\xmg,\xmqu) \biggr] \\
& & \\
S &=& 3\biggl[ \xmqu C_{22}(\xmg,\xmQ,\xmg)-\xmqu C_{23}(\xmg,\xmQ,\xmg)\mp
\mmqu\mmg\sin{(2\xthQ)}C_{12}(\xmg,\xmQ,\xmg)\biggr] \\
& & \mbox{} -\frac{1}{3}\biggl[ -\xmqu C_{22}(\xmQ,\xmg,\xmQ)+\xmqu
C_{23}(\xmQ,\xmg,\xmQ)\mp
\mmqu\mmg\sin{(2\xthQ)}[C_{12}(\xmQ,\xmg,\xmQ)\biggr. \\
& & \mbox{ }\biggl. +\frac{1}{2}C_{0}(\xmQ,\xmg,\xmQ)]\biggr] \\
& & \\
A &=& \pm \frac{3}{2s}\cos{(2\xthQ)}\biggl[ \xmqu [C_{21}(\xmg,\xmQ,\xmg) +
C_{22}(\xmg,\xmQ,\xmg)]+(s-2\xmqu)C_{23}(\xmg,\xmQ,\xmg)\biggr. \\
& & \mbox{ } \biggl. +2C_{24}(\xmg,\xmQ,\xmg)+s\,C_{12}(\xmg,\xmQ,\xmg)-\xmg
C_{0}(\xmg,\xmQ,\xmg)-\frac{2}{9}C_{24}(\xmQ,\xmg,\xmQ)\biggr. \\
& & \mbox{ } \biggl. +\frac{8}{9}B_{1}(\xmQ,\xmg,\xmqu)\biggr]
\end{eqnarray*}

The 4-point integrals in the box terms are
\begin{eqnarray*}
D &=&D(\xmg,\xmq,\xmg,\xmst,0,0,\xmt,\xmt,s,\xmt-2p_1\cdot p_3)
\\
D^c&=&D^c(\xmg,\xmq,\xmg,\xmst,0,0,\xmt,\xmt,s,\xmt-2p_2\cdot
p_3) \;,
\end{eqnarray*}
where $p_1\cdot p_3=s(1-\beta z)/4$ and $p_2\cdot p_3=s(1+\beta z)/4$.  The box
and crossed-box terms are summed over each combination of squarks
$\sq{q}_i\sq{t}_j$, where $i,j=1,2$.  The mixing of the squarks is parameterized
as
\[
\acbd = \mp \frac{1}{4}\sin{(2\xth)} \;,
\]
for $\sq{t}_1$ and $\sq{t}_2$ respectively; and for $\sq{q}_i\sq{t}_j$
\begin{eqnarray*}
\asbs &=&\frac{1}{4}[\cos^2(\xth-\xthq)+\cos^2(\xth+\xthq)] \;, i=j \\
\csds &=&\frac{1}{4}[\sin^2(\xth-\xthq)+\sin^2(\xth+\xthq)] \;, i=j \\
\asbs &=&\frac{1}{4}[\sin^2(\xth-\xthq)+\sin^2(\xth+\xthq)] \;, i\neq j \\
\csds &=&\frac{1}{4}[\cos^2(\xth-\xthq)+\cos^2(\xth+\xthq)] \;, i\neq j \;.
\end{eqnarray*}
\begin{eqnarray*}
B &=& \mmt \mmg s^2(\acbd)[D_{11}-D_{12}+D_{13}+D_{0}]+\xmt
s^2(\asbs)[-D_{12}-D_{23}-D_{24} \\ & & \mbox{ } +D_{26}-2D_{27}/s] +4s(p_2\cdot
p_3)^2(\asbs) [-D_{12}+D_{13}-D_{24}+D_{25}-2D_{27}/s] \\ & & \mbox{ }
+[\xmt s+4(p_1\cdot p_3)^2][\xmt (\asbs)D_{23}+\xmg (\csds)D_{0}-2\mmt
\mmg(\acbd)D_{13}] \\
& & \\
C &=& \mmt \mmg s^2(\acbd)[D^c_{11}-D^c_{12}+D^c_{13}+D^c_{0}]+\xmt
s^2(\csds)[-D^c_{12}-D^c_{23}-D^c_{24} \\ & & \mbox{ } +D^c_{26}-2D^c_{27}/s]
+4s(p_1\cdot p_3)^2(\csds) [-D^c_{12}+D^c_{13}-D^c_{24}+D^c_{25}-2D^c_{27}/s] \\
& & \mbox{ } +[\xmt s+4(p_2\cdot p_3)^2][\xmt (\csds)D^c_{23}+\xmg
(\asbs)D^c_{0}-2\mmt \mmg(\acbd)D^c_{13}]
\end{eqnarray*}

A few equations in the Appendix of Ref.~\cite{ALAM} appear to be misprinted.  As
written they lead to divergent behavior that does not match Fig.~8 in that
paper.  With the following replacements, our analytic results agree up to the
sign discrepency discussed in Sec.~II.
\begin{eqnarray*}
F_{12}^{\rm DB}&=&\frac{\alpha_s}{\pi}\{-\hat{s}(2A_5^{\dagger}A_5^{\dagger})[2
\hat{s}\xmt +2(\hat{u}-\xmt)^2] -\mmg
\mmt(2A_5^{\dagger}A_{5x})[2\hat{s}^2]\}D_{12} \\
F_{13}^{\rm DB}&=&\frac{\alpha_s}{\pi}\{\hat{s}(2A_5^{\dagger}A_5^{\dagger})
[2(\hat{u}-\xmt)^2] +\mmg \mmt(2A_5^{\dagger}A_{5x})[2\hat{s}(\hat{s}-2\xmt)
-4(\hat{t}-\xmt)^2]\}D_{13} \\
F_{12}^{\rm CB}&=&\frac{\alpha_s}{\pi}\{-\hat{s}(2A_5^{\dagger}A_5^{\dagger})[2
\hat{s}\xmt +2(\hat{t}-\xmt)^2] -\mmg
\mmt(2\overline{A_5^{\dagger}A_{5x}})[2\hat{s}^2]\}D_{12} \\
F_{13}^{\rm CB}&=&\frac{\alpha_s}{\pi}\{\hat{s}(2A_5^{\dagger}A_5^{\dagger})
[2(\hat{t}-\xmt)^2] +\mmg \mmt(2\overline{A_5^{\dagger}A_{5x}})[2\hat{s}(\hat{s}
-2\xmt)-4(\hat{u}-\xmt)^2]\}D_{13}
\end{eqnarray*}

\newpage

\begin{center}
\section*{Figure Captions}
\end{center}

Fig. 1:  Feynman diagrams for the one-loop SUSY QCD correction to top
quark production at the Tevatron.  The first row contains the tree-level
diagram.  The second row contains the vacuum polarization correction to the
gluon propagator due to squarks and gluinos.  The third row contains the final
state vertex correction and wave-function renormalization diagrams.  The fourth
row contains the initial state vertex correction and wave-function
renormalization diagrams.  The last row contains the box and crossed-box
diagrams.

Fig. 2:  Contribution of each term to the correction for $p\bar p \to
t\bar t$ as a function of $\mmq =\mmst$, for $\mmg = 200$~GeV.

Fig. 3:  Change in the cross section for $p\bar p \to t\bar t$, as a
function of gluino mass $\mmg$, for $\mmt = 175$~GeV.  Curves of
constant degenerate squark mass $\mmq = \mmst$ are shown.

Fig. 4:  Change in the cross section for $p\bar p \to t\bar t$, as a
function of degenerate squark mass $\mmq = \mmst$, for $\mmt = 175$~GeV.
Curves of constant gluino mass $\mmg$ are shown.

Fig. 5:  Change in the cross section for $p\bar p \to t\bar t$, as a
function of heavy-squark mass $\mmq$, for $\mmt = 175$~GeV, and $\mmg =
200$~GeV.  Curves of constant top-squark mass $\mmst$ are shown.

Fig. 6:  The relative change to the correction is shown as a function of
top-squark mass difference $\Delta\mmst=(\mmb-\mma)$, for various $\mma$, and
mixing angles $\xth$, with $\mmt = 175$~GeV, and $\mmg = 200$~GeV.

Fig. 7:  Differential cross section for $p\bar p \to t\bar t$, as a
function of $t\bar t$ invariant mass $M_{t\bar t}$, for $\mmt = 175$~GeV.
Figures are shown for $\mmg = 150$, 175, 200, and 225~GeV.  Curves of
constant degenerate squark mass $\mmq =\mmst$ are shown.

\newpage

\begin{figure}[tb]
\begin{center}
\epsfxsize= 3.125in
\leavevmode
\epsfbox{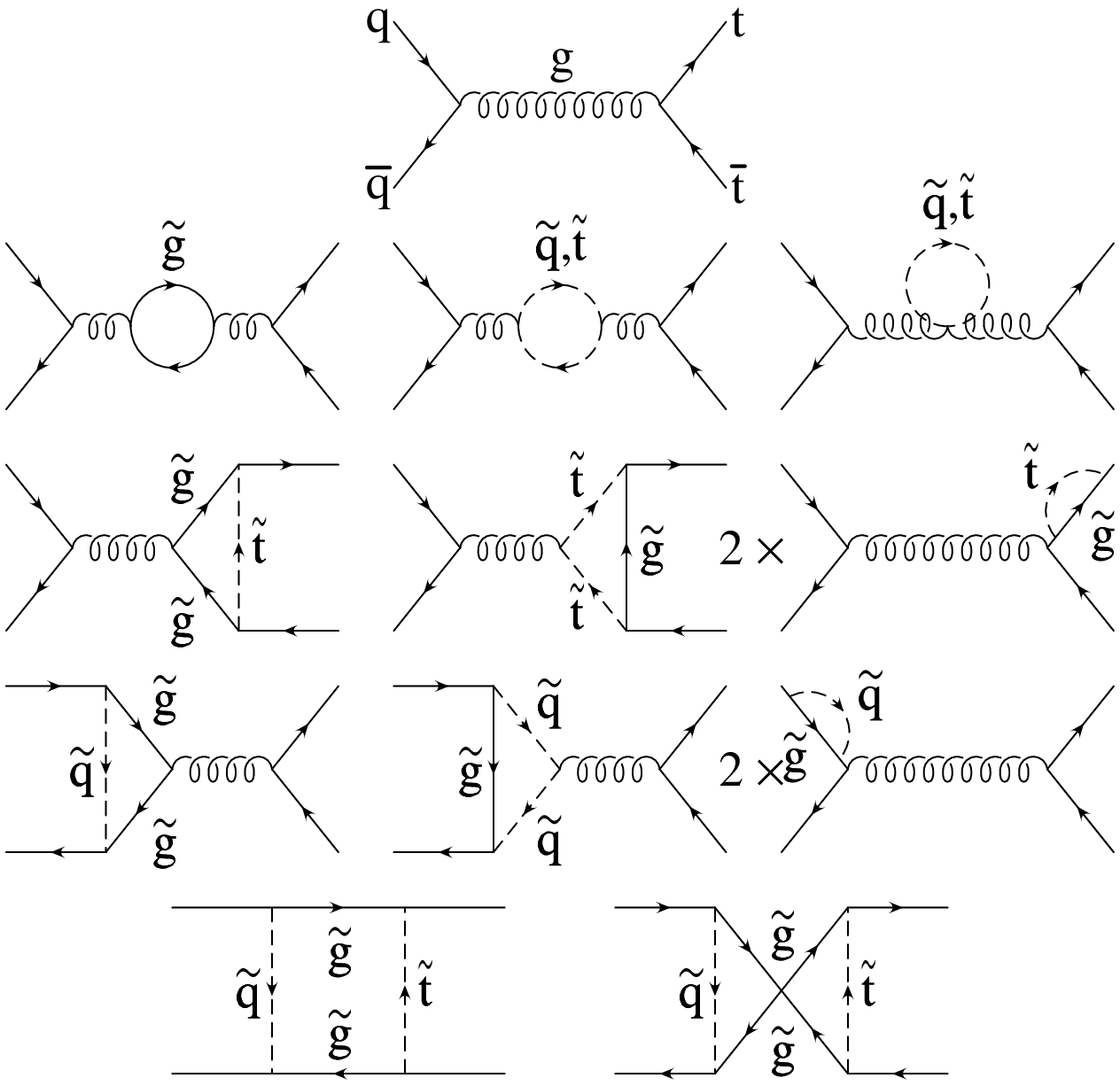} 
\end{center}
\caption[fake]{}
\end{figure}

\begin{figure}[tb]
\begin{center}
\include{range0t}
\end{center}
\caption[fake]{}
\end{figure}

\begin{figure}[tb]
\begin{center}
\include{nrrtt}
\end{center}
\caption[fake]{}
\end{figure}

\begin{figure}[tb]
\begin{center}
\include{rnntt}
\end{center}
\caption[fake]{}
\end{figure}

\begin{figure}[tb]
\begin{center}
\include{20rntt}
\end{center}
\caption[fake]{}
\end{figure}

\begin{figure}[tb]
\begin{center}
\include{allangt}
\end{center}
\caption[fake]{}
\end{figure}

\begin{figure}[tb]
\begin{center}
\include{zrrs}
\end{center}
\caption[fake]{}
\end{figure}

\end{document}

%% file: range0t.tex
\setlength{\unitlength}{0.1bp}
\begin{picture}(2448,1900)(0,0)
\special{psfile=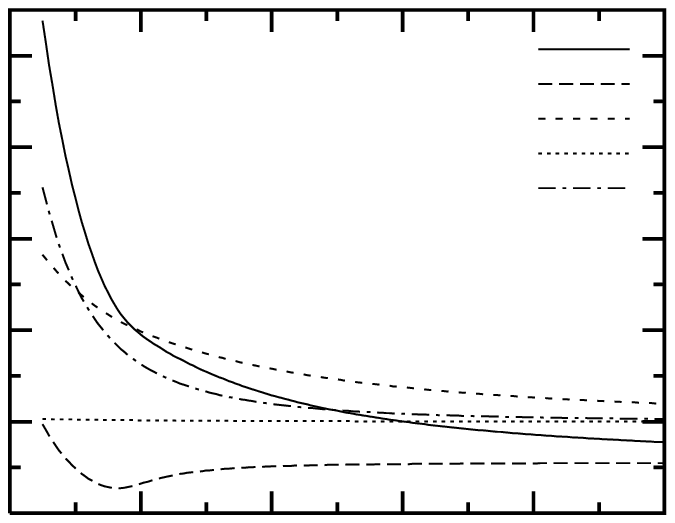 llx=0 lly=0 urx=490 ury=444 rwi=4900}
\put(1872,1287){\makebox(0,0)[r]{Boxes}}
\put(1872,1387){\makebox(0,0)[r]{Scalar}}
\put(1872,1487){\makebox(0,0)[r]{Vector}}
\put(1872,1587){\makebox(0,0)[r]{Vacuum}}
\put(1872,1687){\makebox(0,0)[r]{Total}}
\put(1342,101){\makebox(0,0){$m_{\widetilde{q}}=m_{\widetilde{t}}$~(GeV)}}
\put(150,1075){%
\special{ps: gsave currentpoint currentpoint translate
270 rotate neg exch neg exch translate}%
\makebox(0,0)[b]{\shortstack{$\Delta\sigma/\sigma_0$~(\%)}}%
\special{ps: currentpoint grestore moveto}%
}
\put(2285,251){\makebox(0,0){1000}}
\put(1908,251){\makebox(0,0){800}}
\put(1531,251){\makebox(0,0){600}}
\put(1154,251){\makebox(0,0){400}}
\put(777,251){\makebox(0,0){200}}
\put(400,251){\makebox(0,0){0}}
\put(350,1668){\makebox(0,0)[r]{40}}
\put(350,1405){\makebox(0,0)[r]{30}}
\put(350,1141){\makebox(0,0)[r]{20}}
\put(350,878){\makebox(0,0)[r]{10}}
\put(350,614){\makebox(0,0)[r]{0}}
\put(350,351){\makebox(0,0)[r]{-10}}
\end{picture}

%% file: nrrtt.tex
\setlength{\unitlength}{0.1bp}
\begin{picture}(2448,1900)(0,0)
\special{psfile=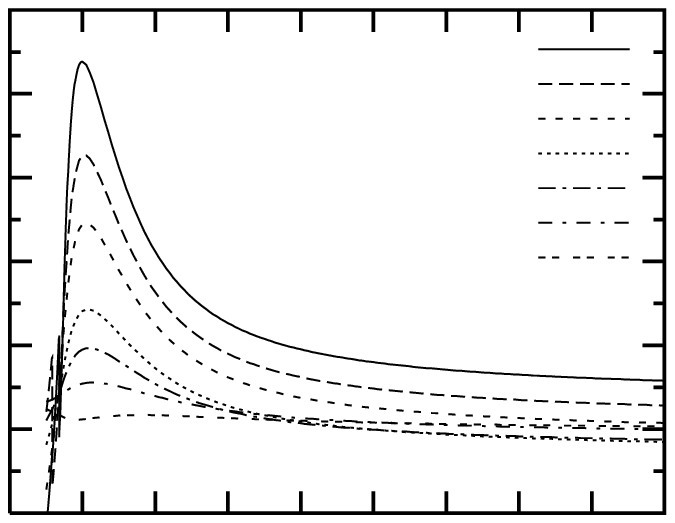 llx=0 lly=0 urx=490 ury=444 rwi=4900}
\put(1872,1087){\makebox(0,0)[r]{500~GeV}}
\put(1872,1187){\makebox(0,0)[r]{300~GeV}}
\put(1872,1287){\makebox(0,0)[r]{200~GeV}}
\put(1872,1387){\makebox(0,0)[r]{150~GeV}}
\put(1872,1487){\makebox(0,0)[r]{100~GeV}}
\put(1872,1587){\makebox(0,0)[r]{75~GeV}}
\put(1872,1662){\makebox(0,0)[r]{$m_{\widetilde{t}}=m_{\widetilde{q}}=50$~GeV}}
\put(1342,101){\makebox(0,0){$m_{\widetilde{g}}$~(GeV)}}
\put(150,1075){%
\special{ps: gsave currentpoint currentpoint translate
270 rotate neg exch neg exch translate}%
\makebox(0,0)[b]{\shortstack{$\Delta\sigma /\sigma_0$~(\%)}}%
\special{ps: currentpoint grestore moveto}%
}
\put(2285,251){\makebox(0,0){1000}}
\put(2076,251){\makebox(0,0){900}}
\put(1866,251){\makebox(0,0){800}}
\put(1657,251){\makebox(0,0){700}}
\put(1447,251){\makebox(0,0){600}}
\put(1238,251){\makebox(0,0){500}}
\put(1028,251){\makebox(0,0){400}}
\put(819,251){\makebox(0,0){300}}
\put(609,251){\makebox(0,0){200}}
\put(400,251){\makebox(0,0){100}}
\put(350,1800){\makebox(0,0)[r]{50}}
\put(350,1559){\makebox(0,0)[r]{40}}
\put(350,1317){\makebox(0,0)[r]{30}}
\put(350,1076){\makebox(0,0)[r]{20}}
\put(350,834){\makebox(0,0)[r]{10}}
\put(350,593){\makebox(0,0)[r]{0}}
\put(350,351){\makebox(0,0)[r]{-10}}
\end{picture}

%% file: rnntt.tex
\setlength{\unitlength}{0.1bp}
\begin{picture}(2448,1900)(0,0)
\special{psfile=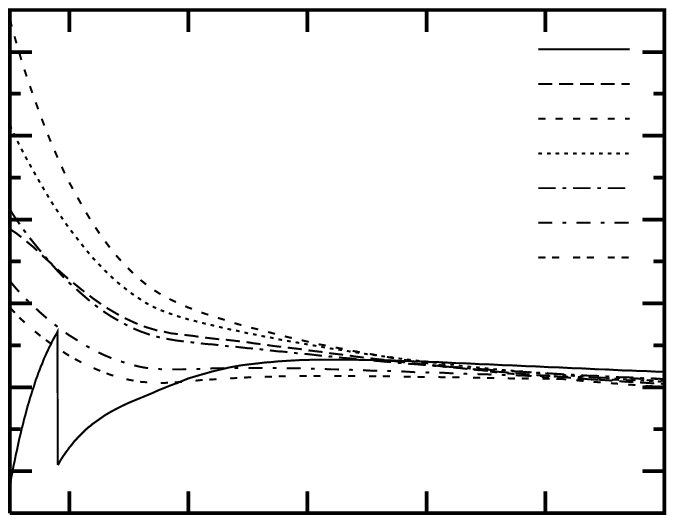 llx=0 lly=0 urx=490 ury=444 rwi=4900}
\put(1872,1087){\makebox(0,0)[r]{500~GeV}}
\put(1872,1187){\makebox(0,0)[r]{400~GeV}}
\put(1872,1287){\makebox(0,0)[r]{300~GeV}}
\put(1872,1387){\makebox(0,0)[r]{250~GeV}}
\put(1872,1487){\makebox(0,0)[r]{200~GeV}}
\put(1872,1587){\makebox(0,0)[r]{175~GeV}}
\put(1872,1662){\makebox(0,0)[r]{$m_{\widetilde{g}}=150$~GeV}}
\put(1342,101){\makebox(0,0){$m_{\widetilde{q}}=m_{\widetilde{t}}$~(GeV)}}
\put(150,1075){%
\special{ps: gsave currentpoint currentpoint translate
270 rotate neg exch neg exch translate}%
\makebox(0,0)[b]{\shortstack{$\Delta\sigma /\sigma_0$~(\%)}}%
\special{ps: currentpoint grestore moveto}%
}
\put(2285,251){\makebox(0,0){600}}
\put(1942,251){\makebox(0,0){500}}
\put(1600,251){\makebox(0,0){400}}
\put(1257,251){\makebox(0,0){300}}
\put(914,251){\makebox(0,0){200}}
\put(571,251){\makebox(0,0){100}}
\put(350,1679){\makebox(0,0)[r]{40}}
\put(350,1438){\makebox(0,0)[r]{30}}
\put(350,1196){\makebox(0,0)[r]{20}}
\put(350,955){\makebox(0,0)[r]{10}}
\put(350,713){\makebox(0,0)[r]{0}}
\put(350,472){\makebox(0,0)[r]{-10}}
\end{picture}

%% file: 20rntt.tex
\setlength{\unitlength}{0.1bp}
\begin{picture}(2448,1900)(0,0)
\special{psfile=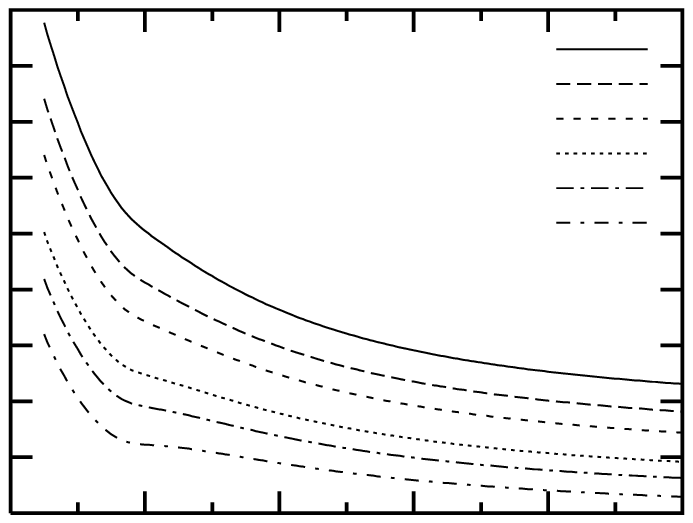 llx=0 lly=0 urx=490 ury=444 rwi=4900}
\put(1872,1187){\makebox(0,0)[r]{300~GeV}}
\put(1872,1287){\makebox(0,0)[r]{200~GeV}}
\put(1872,1387){\makebox(0,0)[r]{150~GeV}}
\put(1872,1487){\makebox(0,0)[r]{100~GeV}}
\put(1872,1587){\makebox(0,0)[r]{75~GeV}}
\put(1872,1662){\makebox(0,0)[r]{$m_{\widetilde{t}}=50$~GeV}}
\put(1317,101){\makebox(0,0){$m_{\widetilde{q}}$~(GeV)}}
\put(150,1075){%
\special{ps: gsave currentpoint currentpoint translate
270 rotate neg exch neg exch translate}%
\makebox(0,0)[b]{\shortstack{$\Delta\sigma /\sigma_0$~(\%)}}%
\special{ps: currentpoint grestore moveto}%
}
\put(2285,251){\makebox(0,0){1000}}
\put(1898,251){\makebox(0,0){800}}
\put(1511,251){\makebox(0,0){600}}
\put(1124,251){\makebox(0,0){400}}
\put(737,251){\makebox(0,0){200}}
\put(350,251){\makebox(0,0){0}}
\put(300,1800){\makebox(0,0)[r]{45}}
\put(300,1639){\makebox(0,0)[r]{40}}
\put(300,1478){\makebox(0,0)[r]{35}}
\put(300,1317){\makebox(0,0)[r]{30}}
\put(300,1156){\makebox(0,0)[r]{25}}
\put(300,995){\makebox(0,0)[r]{20}}
\put(300,834){\makebox(0,0)[r]{15}}
\put(300,673){\makebox(0,0)[r]{10}}
\put(300,512){\makebox(0,0)[r]{5}}
\put(300,351){\makebox(0,0)[r]{0}}
\end{picture}

%% file: allangt.tex
\setlength{\unitlength}{0.1bp}
\begin{picture}(2448,2483)(0,0)
\special{psfile=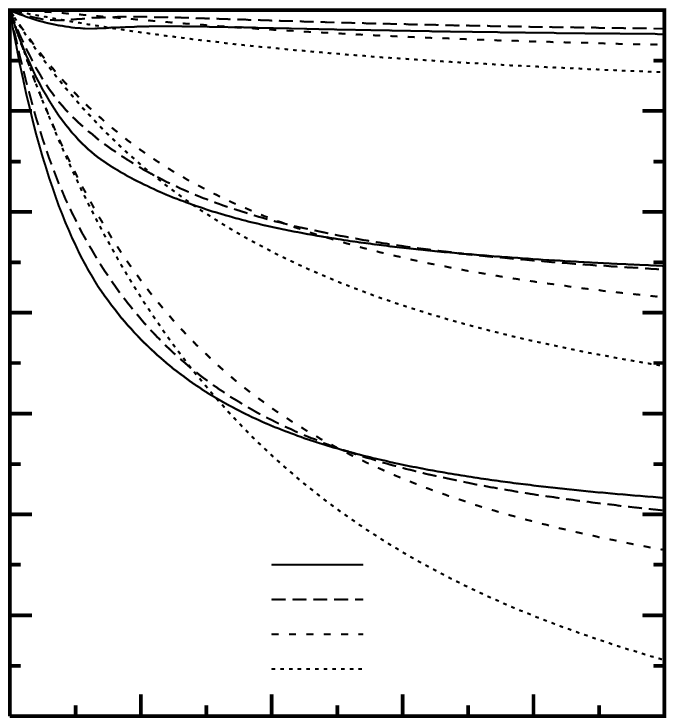 llx=0 lly=0 urx=490 ury=580 rwi=4900}
\put(1104,487){\makebox(0,0)[r]{500~GeV}}
\put(1104,587){\makebox(0,0)[r]{300~GeV}}
\put(1104,687){\makebox(0,0)[r]{100~GeV}}
\put(1104,762){\makebox(0,0)[r]{$m_{\widetilde{t}_1} = 50$~GeV}}
\put(1720,1121){\makebox(0,0)[l]{$\theta_{\widetilde{t}} = 45^\circ$}}
\put(1720,1760){\makebox(0,0)[l]{$\theta_{\widetilde{t}} = 90^\circ$}}
\put(1720,2137){\makebox(0,0)[l]{$\theta_{\widetilde{t}} = 135^\circ$}}
\put(1342,101){\makebox(0,0){$\Delta m_{\widetilde{t}}$~(GeV)}}
\put(150,1367){%
\special{ps: gsave currentpoint currentpoint translate
270 rotate neg exch neg exch translate}%
\makebox(0,0)[b]{\shortstack{$\Delta\sigma(\Delta m_{\widetilde{t}})/\Delta\sigma(0)$}}%
\special{ps: currentpoint grestore moveto}%
}
\put(2285,251){\makebox(0,0){1000}}
\put(1908,251){\makebox(0,0){800}}
\put(1531,251){\makebox(0,0){600}}
\put(1154,251){\makebox(0,0){400}}
\put(777,251){\makebox(0,0){200}}
\put(400,251){\makebox(0,0){0}}
\put(350,2384){\makebox(0,0)[r]{1.0}}
\put(350,2094){\makebox(0,0)[r]{0.8}}
\put(350,1803){\makebox(0,0)[r]{0.6}}
\put(350,1513){\makebox(0,0)[r]{0.4}}
\put(350,1222){\makebox(0,0)[r]{0.2}}
\put(350,932){\makebox(0,0)[r]{0}}
\put(350,641){\makebox(0,0)[r]{-0.2}}
\put(350,351){\makebox(0,0)[r]{-0.4}}
\end{picture}

%% file: zrrs.tex
\setlength{\unitlength}{0.1bp}
\begin{picture}(2448,1468)(0,0)
\special{psfile=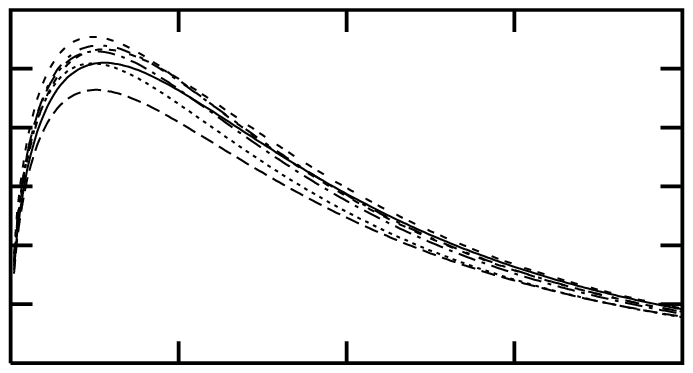 llx=0 lly=0 urx=490 ury=343 rwi=4900}
\put(150,859){%
\special{ps: gsave currentpoint currentpoint translate
270 rotate neg exch neg exch translate}%
\makebox(0,0)[b]{\shortstack{$d\sigma /dM_{t\bar t}$~(fb/GeV)}}%
\special{ps: currentpoint grestore moveto}%
}
\put(800,600){\makebox(0,0){$m_{\widetilde{g}}=150$~GeV}}
\put(300,1368){\makebox(0,0)[r]{30}}
\put(300,1199){\makebox(0,0)[r]{25}}
\put(300,1029){\makebox(0,0)[r]{20}}
\put(300,860){\makebox(0,0)[r]{15}}
\put(300,690){\makebox(0,0)[r]{10}}
\put(300,521){\makebox(0,0)[r]{5}}
\put(300,351){\makebox(0,0)[r]{0}}
\end{picture}
\begin{picture}(2448,1018)(0,0)
\special{psfile=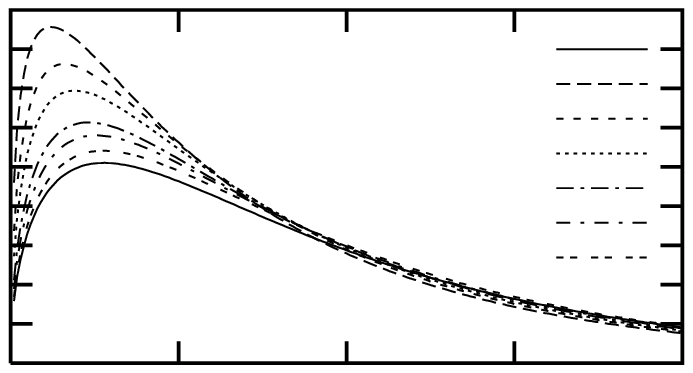 llx=0 lly=0 urx=490 ury=343 rwi=4900}
\put(1872,655){\makebox(0,0)[r]{300~GeV}}
\put(1872,755){\makebox(0,0)[r]{200~GeV}}
\put(1872,855){\makebox(0,0)[r]{150~GeV}}
\put(1872,955){\makebox(0,0)[r]{100~GeV}}
\put(1872,1055){\makebox(0,0)[r]{75~GeV}}
\put(1872,1130){\makebox(0,0)[r]{$m_{\widetilde{t}}=m_{\widetilde{q}}=50$~GeV}}
\put(1872,1255){\makebox(0,0)[r]{Tree Level}}
\put(150,859){%
\special{ps: gsave currentpoint currentpoint translate
270 rotate neg exch neg exch translate}%
\makebox(0,0)[b]{\shortstack{$d\sigma /dM_{t\bar t}$~(fb/GeV)}}%
\special{ps: currentpoint grestore moveto}%
}
\put(800,600){\makebox(0,0){$m_{\widetilde{g}}=175$~GeV}}
\put(300,1255){\makebox(0,0)[r]{40}}
\put(300,1142){\makebox(0,0)[r]{35}}
\put(300,1029){\makebox(0,0)[r]{30}}
\put(300,916){\makebox(0,0)[r]{25}}
\put(300,803){\makebox(0,0)[r]{20}}
\put(300,690){\makebox(0,0)[r]{15}}
\put(300,577){\makebox(0,0)[r]{10}}
\put(300,464){\makebox(0,0)[r]{5}}
\put(300,351){\makebox(0,0)[r]{0}}
\end{picture}
\begin{picture}(2448,1018)(0,0)
\special{psfile=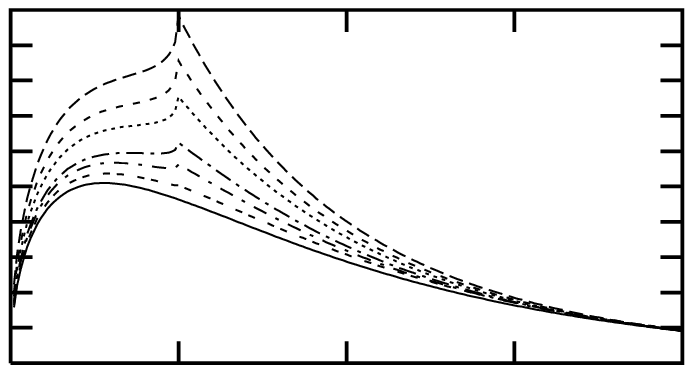 llx=0 lly=0 urx=490 ury=343 rwi=4900}
\put(150,859){%
\special{ps: gsave currentpoint currentpoint translate
270 rotate neg exch neg exch translate}%
\makebox(0,0)[b]{\shortstack{$d\sigma /dM_{t\bar t}$~(fb/GeV)}}%
\special{ps: currentpoint grestore moveto}%
}
\put(800,600){\makebox(0,0){$m_{\widetilde{g}}=200$~GeV}}
\put(300,1266){\makebox(0,0)[r]{45}}
\put(300,1165){\makebox(0,0)[r]{40}}
\put(300,1063){\makebox(0,0)[r]{35}}
\put(300,961){\makebox(0,0)[r]{30}}
\put(300,860){\makebox(0,0)[r]{25}}
\put(300,758){\makebox(0,0)[r]{20}}
\put(300,656){\makebox(0,0)[r]{15}}
\put(300,554){\makebox(0,0)[r]{10}}
\put(300,453){\makebox(0,0)[r]{5}}
\put(300,351){\makebox(0,0)[r]{0}}
\end{picture}
\begin{picture}(2448,1018)(0,0)
\special{psfile=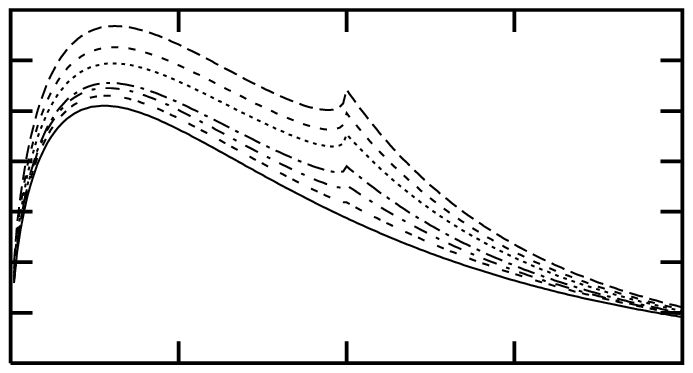 llx=0 lly=0 urx=490 ury=343 rwi=4900}
\put(1317,101){\makebox(0,0){$M_{t\bar t}$~(GeV)}}
\put(150,859){%
\special{ps: gsave currentpoint currentpoint translate
270 rotate neg exch neg exch translate}%
\makebox(0,0)[b]{\shortstack{$d\sigma /dM_{t\bar t}$~(fb/GeV)}}%
\special{ps: currentpoint grestore moveto}%
}
\put(800,600){\makebox(0,0){$m_{\widetilde{g}}=225$~GeV}}
\put(2285,251){\makebox(0,0){550}}
\put(1801,251){\makebox(0,0){500}}
\put(1318,251){\makebox(0,0){450}}
\put(834,251){\makebox(0,0){400}}
\put(350,251){\makebox(0,0){350}}
\put(300,1223){\makebox(0,0)[r]{30}}
\put(300,1077){\makebox(0,0)[r]{25}}
\put(300,932){\makebox(0,0)[r]{20}}
\put(300,787){\makebox(0,0)[r]{15}}
\put(300,642){\makebox(0,0)[r]{10}}
\put(300,496){\makebox(0,0)[r]{5}}
\put(300,351){\makebox(0,0)[r]{0}}
\end{picture}

%% file: ttsusy.bbl
\newpage

\begin{thebibliography}{99}

\bibitem{TOPD} F.~Abe~{\it et~al.}, CDF~Collaboration, \PRL 74 2626
1995 ; S.~Abachi~{\it et~al.}, D0~Collaboration, \PRL 74 2632 1995 .

\bibitem{TOPM} D.~Gerdes, CDF~Collaboration, FERMILAB-CONF-96/342-E. Proceedings
1996 DPF/DPB Summer Study, ``New Directions for High Energy Physics,'' Snowmass,
CO, June 25-July 12, 1996.

\bibitem{TEV2000} ``Future of Electroweak Physics at the Fermilab Tevatron:
Report of the tev\_2000 Study Group'', edited by D.~Amidei and R.~Brock,
FERMILAB-PUB-96/082, 1996.

\bibitem{NILLESAR} For reviews, see H.~P.~Nilles, \PREP 110 1 1984 ; P.~Nath,
R.~Arnowitt, and A.~Chamseddine, {\it Applied N=1 Supergravity}, ICTP series in
Theoretical Physics, (World Scientific, Singapore, 1984); Also see references
in J.~Amundson {\it et~al.}, hep-ph/9609374.  Proceedings of the 1996 DPF/DPB
Summer Study, ``New Directions for High Energy Physics,'' Snowmass, CO, June
25-July 12, 1996.

\bibitem{ELLISMIX} J.~Ellis and S.~Rudaz, \PLB 128 248 1983 ; A.~Bouquet,
J.~Kaplan and C.~Savoy, \NPB 262 299 1985 .

\bibitem{ALSUSY} M. Schmitt (private communication), ALEPH preliminary results
presented at the ``LEP Jamboree'', October 8, 1996.

\bibitem{CDFSG} F.~Abe~{\it et~al.}, CDF~Collaboration, \PRL 76 2006 1996 .

\bibitem{DATTA} A.~Datta, M.~Guchait, and N.~Parua, hep-ph/9609413 and
references therein.

\bibitem{OSUSY} S.~Abachi {\it et~al.}, D0~Collaboration,
FERMILAB-Conf-95/193-E, D0~Note~2614, 1995; H1~Collaboration, hep-ex/9605002.

\bibitem{D0ST} S.~Abachi {\it et~al.}, D0~Collaboration, \PRL 76 2222 1996 .

\bibitem{CROSSTH} P.~Nason, S.~Dawson, and R.~K.~Ellis, \NPB 303 607 1988 ; \NPB
327 49 1989 ; W.~Beenakker, H.~Kuijif, W.~van~Neerven, and J.~Smith, \PRD 40 54
1989 ; W.~Beenakker, W.~van~Neerven, R.~Meng, G.~Schuler, and J.~Smith, \NPB 351
507 1991 ; E.~Laenen, J.~Smith, and W.~van~Neerven, \NPB 369 543 1992 ; \PLB 321
254 1994 ; S.~Catani, M.~Mangano, P.~Nason, and L.~Trentadue, \PLB 378 329 1996
; hep-ph/9604351; E.~Berger and H.~Contopanagos, \PLB 361 115 1995 ; \PRD 54
3085 1996 .

\bibitem{DJOUADI} A.~Djouadi, M.~Drees, and H.~K\"{o}nig, \PRD 48 3081 1993 .

\bibitem{LIHU} C.~Li, B.~Hu, J.~Yang, and C.~Hu, \PRD 52 5014 1995 ; Erratum:
\PRD 53 4112 1996 .

\bibitem{KIMLOP} J.~Kim, J.~Lopez, D.~V.~Nanopoulos, and R.~Rangarajan, \PRD 54
4364 1996 .

\bibitem{ALAM} S.~Alam, K.~Hagiwara, and S.~Matsumoto, \PRD 55 1307 1997 .

\bibitem{HABERK} H.~Haber and G.~Kane, \PREP 117 75 1985 .

\bibitem{PASSARINOV} G.~Passarino and M.~Veltman, \NPB 160 151 1979 .

\bibitem{FF} G.~J.~van Oldenborgh, Comput.~Phys.~Commun.~{\bf 66}, 1 (1991).

\bibitem{MRSAP} A.~D.~Martin, W.~J.~Stirling, and R.~G.~Roberts, \PRD 51 4756
1995 .

\bibitem{CTEQ} H.~Lai, J.~Botts, J.~Huston, J.~Morfin, J.~Owens, 
J.~Qiu, W.-K.~Tung, and H.~Weerts, \PRD 51 4763 1995 .

\bibitem{THOOFTV} G.~'t~Hooft and M.~Veltman, \NPB 153 365 1979 .

\bibitem{Itzykson} C.~Itzykson and J.~Zuber, {\it Quantum Field Theory}
(McGraw-Hill, New York, 1980), p. 311.

\bibitem{BIGI} I.~Bigi, Y.~Dokshitzer, V.~Khoze, J.~K\"{u}hn, and P.~Zerwas,
\PLB 181 157 1986 .

\bibitem{JEZ} M.~Je\.{z}abek, \NPBPS 37 197 1994 .

\bibitem{ELLISR} J.~Ellis and D.~Ross, \PLB 383 187 1996 .

\end{thebibliography}
